\documentclass[twocolumn,prl]{revtex4-1} 
\def\Title {Josephson relation for disordered superfluids}

\usepackage[pdftex]{hyperref}	
\usepackage{amsmath,amsfonts,amssymb,bm}

\usepackage[pdftex]{graphicx}

\newcommand{\be}{\begin{equation}}
\newcommand{\ee}{\end{equation}}
\newcommand{\nc}{n_\text{c}}
\newcommand{\ncoh}{n_\text{coh}}

\newcommand{\vc}[1]{\bm{#1}}
\newcommand{\br}{{\vc r}}
\newcommand{\bk}{{\vc k}}\newcommand{\bj}{{\vc j}}
\newcommand{\bq}{{\vc q}}
\newcommand{\bp}{{\vc p}}

\newcommand{\bnabla}{\boldsymbol\nabla}

\newcommand{\rmd}{\mathrm{d}}

\newcommand{\epn}[1]{\epsilon^0_{#1}}		
\newcommand{\ep}[1]{\epsilon_{#1}}		

\newcommand{\avg}[1]{\overline{{#1}}}	

\newcommand{\xpct}[1]{\bigl\langle #1 \bigr\rangle} 

\newcommand{\rhos}{\rho_\text{s}}
\newcommand{\rhoc}{\rho_\text{c}}
\newcommand{\rhon}{\rho_\text{n}}

\begin{document}

\title{\Title}
\author{Cord A.\ M\"uller}
\affiliation
{Fachbereich Physik, Universit\"at Konstanz, 78457 Konstanz, Germany}
\affiliation{Institut Non Lin\'eaire de Nice, CNRS and Universit\'e Nice--Sophia Antipolis, 06560 Valbonne, France}
\date{\today} 
\begin{abstract} 
The Josephson sum rule relates the superfluid density to the
condensate order parameter, via the infrared residue of the
single-particle Green's function. We establish an effective Josephson
relation for disordered condensates valid upon ensemble
averaging. This relation has the merit to show explicitly how
superfluidity links to the coherent density, i.e., 
the density of particles with zero momentum. 
Detailed agreement is reached with perturbation theory for weak disorder.
\end{abstract}

\maketitle

\paragraph{Introduction:---} 
Bose-Einstein condensation (BEC) and superfluidity are certainly linked,
and yet this link is difficult to state with precision in situations that involve, e.g., strong interactions, 
low dimensions, external potentials or temperatures close to critical. Josephson
\cite{Josephson1966} has derived a relation between the superfluid mass
density $\rhos$ and the BEC order parameter $\psi$ that provides such
a link: 
\be\label{cleanJoseph}
 \frac{m|\psi|^2}{\rhos}
=-\lim_{\bk\to0}\frac{\bk^2}{m}G(\bk,0). 
\ee
Here, $m$ is the mass of the individual bosonic particle, and
$G(\bk,0)$ is the single-particle Green's function at momentum $\bk$
and zero (Matsubara) frequency. Because the Green
function can be written as a frequency integral 
over its imaginary part, the spectral function, this relation is also referred to as the Josephson
sum rule \cite{Baym1969,Ueda2010}. Only within mean-field theory, neglecting quantum and thermal fluctuations, one finds that $\rhos=m|\psi|^2$ (see
eq.~\eqref{mfclean} below), and there
is no need for subtle distinctions between condensate and
superfluid. But especially under critical conditions, 
the Josephson relation is precious because it connects the scaling properties of
condensate and superfluid order parameters through the Josephson (hyper-)scaling law \cite{Josephson1966,Holzmann2003,Holzmann2007}. 

Because of its conceptual and practical importance, the Josephson relation has
been re-derived over the years using various methods
\cite{Baym1969,Griffin1984,Holzmann2007,Ueda2010,Dawson2012}. These derivations
all make use of translational invariance and thus are only valid, strictly
speaking, in clean systems. Although the Josephson scaling law has been occasionally applied
\cite{Crowell1997,Balibar2008} (and questioned
\cite{Reppy1992}) in disordered systems, it is not immediately clear how
to read the relation \eqref{cleanJoseph} in that case. Indeed, the BEC
order parameter
$\psi(\br)$ acquires a spatial dependence on each realization of
disorder, and also the Green function
is no longer diagonal in momentum. Since one can take $\rhos$ to be a self-averaging 
quantity in a bulk system of linear size $L$, one may be tempted to think  
that \eqref{cleanJoseph}  should hold under the ensemble average, noted by the overbar $\avg{(\dots)}$: 
\be\label{disorderedJoseph1}
\frac{m\avg{|\psi|^2}}{\rhos}
= -\lim_{\bk\to0}\frac{\bk^2}{m}\avg{G(\bk,0)}. 
\ee
If this were true, the Josephson relation would constrain the ratio of superfluid density to the average
condensate density 
\footnote{We assume statistical homogeneity and trade ensemble for spatial averages where appropriate.}, 
\be\label{nc}
\avg{|\psi|^2} = L^{-d}\int\rmd \br
|\psi(\br)|^2=:\nc.
\ee
The purpose of this paper is to show that this is \emph{not} the case. In the following, the correct Josephson relation is first stated and briefly discussed, then derived, and finally analytically checked in the simplest accessible regime of low temperatures, weak interactions, and weak disorder.

\paragraph{Inhomogeneous Josephson relation:---}
Our main result is the following  Josephson relation for inhomogeneous
systems valid upon ensemble averaging:  
\be\label{joseph}
\frac{m\left|\avg{\psi}\right|^2}{\rhos} = -\lim_{\bk\to0}\frac{\bk^2}{m}
\avg{G(\bk,0)}.  
\ee  
Here, instead of the average condensate density \eqref{nc}, it is the \emph{coherent
density} 
\be
\left|\avg{\psi}\right|^2 =
\left|L^{-d}\int\rmd\br\psi(\br)\right|^2=:\ncoh, 
\ee
of condensed particles 
with $\bk=0$, 
 that is linked with the peculiar long-range,
phase-coherent transport properties that we call superfluid
stiffness. 
The coherent density can be  defined equivalently by $n_\text{coh} =
\lim_{|\br|\to\infty} \avg{\xpct{\hat\psi^\dagger(\br)\hat\psi(0)}}$
as the component with off-diagonal long range order of the
\emph{ensemble-averaged} one-body density matrix. 
As recognized already by Penrose and Onsager \cite{Penrose1956}, in systems that are not fully translation invariant, the
condensate properly speaking comprises all particles
in the maximally populated eigenmode $\psi(\br)$ \cite{Astrakharchik2011,Mueller2012} and 
thus contains the coherent component with $\bk=0$ plus the ``glassy'' component with $\bk\neq0$
\cite{Yukalov2007,Krumnow2011}.

Qualitatively, this strong link between superfluid and coherent density may not surprise much, 
and indeed it has been observed in numerical calculations \cite{Pilati2009,Carleo2013} that 
superfluid and coherent fractions vanish together at (one and the same) superfluid-insulator
critical point, as implied by a finite right-hand side of 
\eqref{joseph} at criticality. The preference of \eqref{joseph} over \eqref{disorderedJoseph1} is also consistent
with the view that the insulating Bose glass close to the transition is a collection of locally
condensed puddles with finite mean density \eqref{nc}, which fail to connect phase-coherently over the full
system size \cite{Yamamoto2008,Falco2009a,Falco2009b,Diallo2014}. However, to our knowledge
a quantitative statement such as \eqref{joseph} has not been put on record 
before. Moreover, recent numerical 
results in $d=2$ \cite{Astrakharchik2013} seem to suggest that superfluid and coherent density do not vanish together. Therefore, we believe it worthwhile to derive \eqref{joseph} by a 
microscopic calculation and check its validity in an analytically tractable limit.

\paragraph{Derivation:---} 

We consider a single-component Bose-condensed fluid with repulsive
interactions in its kinematic ground state,
at inverse temperature $\beta$, confined to a $d$-dimensional volume
of linear size $L$ and subject to an external one-body potential
$V(\br)$.    
 The total average density $n = L^{-d}\int \rmd\br
\xpct{\hat\psi^\dagger(\br)\hat\psi(\br)}$ is fixed by the chemical
potential $\mu$ and splits into the sum of the condensate density
\eqref{nc} and the non-condensed density. The latter comprises quantum depleted and, at $T>0$,  thermally excited particles.  
The condensate is described by a scalar, stationary BEC order parameter $\psi(\br)$. Such an order
parameter may be defined as the macroscopically populated eigenmode of
the one-body density matrix \cite{Penrose1956}.
In the U(1)
symmetry-breaking picture of BEC \cite{Hohenberg1965},  
one rather defines $\psi(\br) = \xpct{\hat\psi(\br)}$ as the expectation
value of the bosonic field operator; 
we use the latter definition for its technical
simplicity. 

Following Baym \cite{Baym1969} (see also \cite{Ueda2010}), we calculate via linear response 
how much adding a particle with momentum $\bk$ changes the order parameter on
the one hand, and the current density on the other.  
We assume that the external potential is an ergodic random process, 
and reach translation invariance by ensemble-averaging. 
Comparing the changes in order parameter and current density then
leads to the generalized Josephson relation \eqref{joseph}.      

To this end, let 
\be\label{deltaH}
\delta\hat H_\bk = \delta \, \hat a_\bk^\dagger = \frac{\delta}{L^{d/2}}
\int\rmd\br e^{-i\bk\cdot\br} \hat\psi^\dagger(\br)
\ee
 be the small perturbation ($|\delta|\ll\mu$) that adds a particle with
 momentum $\bk$ to the system  
\footnote{In a number-conserving
 setting, one could consider $\delta\hat H_\bk =  \delta  \hat
 a_\bk^\dagger \hat a_0$, i.e., the promotion of a particle from zero
 to finite momentum.}.
The  linear response of the condensate amplitude on
 average is   
\be
\avg{\delta\psi(\br)} = - \int_0^\beta\rmd \tau \avg{\xpct{\hat \psi(\br,\tau) \delta\hat H_\bk(0)}}
=  \frac{\delta}{L^{d/2}} e^{i\bk\br} \avg{G(\bk,0)}, 
\ee
which brings about the zero-frequency component of the ensemble-averaged
Matsubara-Green function 
\be
\avg{G(\bk,i\omega_n)} = - \int \rmd\br e^{-i\bk\cdot\br} \int_0^\beta\rmd\tau
e^{i\omega_n\tau}\avg{\xpct{\hat\psi(\br,\tau)\hat\psi^\dagger(0,0)}}. 
\ee
(If the ensemble average were not taken at this stage, one would face a
Green function that is not diagonal in $\bk$, which would compromise
the following derivation.) 
The average condensate amplitude, 
 \be
\avg{\psi(\br)}+\avg{\delta\psi(\br)} =:  \avg{\psi(\br)}\left[1+i\,\delta\varphi\right],  
\ee
is changed by a pure phase factor when 
\be\label{delphi}
\delta \varphi = -i\frac{\avg{\delta\psi(\br)}}{\avg{\psi(\br)}} = \frac{-i\delta}{L^{d/2}\avg{\psi(\br)}}  e^{i\bk\cdot\br} \avg{G(\bk,0)} 
\ee
is real, which can be realized by choosing the phase of $\delta$
appropriately and in the limit $\bk\to0$ (this is the step where taking 
the limit is required).  
This phase's gradient then induces on average the superfluid mass current 
\be\label{deljfrompsi}
m\,\avg{\delta\bj(\br)} 
= \left. \frac{\rhos}{m} \bnabla\delta\varphi \right|_{\bk\to 0} 
=  \delta\frac{  \rhos \bk  e^{i\bk\cdot\br}}{L^{d/2}m \avg{\psi(\br)}} 
\left.
\avg{G(\bk,0)}\right|_{\bk\to 0}. 
\ee

Now we calculate the current directly via linear response,   
\be
\delta \bj(\br) =  - \int_0^\beta\rmd \tau \xpct{\hat \bj(\br,\tau)\delta\hat H_\bk(0)}. 
\ee
Yet, even for a perturbation as simple as \eqref{deltaH}, this is in
general impossible, for one cannot compute the full time-dependence
of the current in the presence of interactions. But 
we can invoke particle number conservation,  as expressed by the continuity
equation, in imaginary time: 
\be\label{continuity} 
i\partial_\tau\hat n(\br,\tau)+\nabla\cdot\hat\bj(\br,\tau)=0. 
\ee 
(Its proof is elementary: Given the Hamiltonian $\hat
H[\hat\psi,\hat\psi^\dagger]=\hat K+ \hat U$ with kinetic energy 
$\hat K = \frac{1}{2m} \int \rmd \br \bnabla\hat \psi^\dagger\cdot\bnabla\hat\psi$,
and an interaction $\hat U=U[\hat n]$ that is a functional of the density
only,  \eqref{continuity} is equivalent to 
the equation of motion 
$\partial_\tau\hat n = [\hat K,\hat n]$.) 
In the momentum representation, the continuity equation \eqref{continuity} becomes 
\be
\partial_\tau \hat n_\bp(\tau) + \bp \cdot\hat\bj_\bp(\tau) = 0,  
\ee
and thus permits to replace the longitudinal current by the density
variation according to   
$ |\bp| \hat j_\bp^\parallel  = - \partial_\tau \hat n_\bp $. 
This allows us to evaluate the Matsubara-time integral,  
\be
\delta j^\parallel_\bp  =   |\bp|^{-1} \int_0^\beta\rmd\tau \xpct{
  \partial_\tau \hat n_\bp(\tau)\delta \hat H_\bk(0) } = 
- |\bp|^{-1} \xpct{[\hat n _\bp,\delta \hat H_\bk]}  
\ee
and we are left with the simple equal-time commutator 
\be 
[\hat n _\bp,\delta \hat H_\bk] = \delta\,  \hat a^\dagger_{\bk-\bp}. 
\ee
Thus we find after ensemble-averaging 
\be
\avg{\delta j^\parallel(\br)} = - \frac{\delta}{L^{d/2}|\bk| } \avg{\psi^*(\br)} e^{i\bk\cdot\br}.    
\ee
Comparing this result with \eqref{deljfrompsi}, whose leading contribution in the limit $\bk\to0$ is also purely longitudinal, then establishes \eqref{joseph}. We remark that the zero-frequency Green function appearing here contains the full dynamical single-particle correlations and can in general not be reduced to the equal-time momentum distribution that enters, for instance, the one-body density matrix \cite{Pitaevskii2003}.

\paragraph{Consistency check in perturbation theory:---} Exact
analytical results are hard to obtain, but we can evaluate the factors entering 
\eqref{joseph} perturbatively for weak disorder using inhomogeneous quadratic
Bogoliubov theory \cite{Gaul2011a,Mueller2012} and check whether they
match. 

First, the  coherent density is given by eq.~(11) in
\cite{Mueller2012},  
\be\label{cohfrac}
\ncoh = \left|\avg{\psi}\right|^2 = \nc [1-V_2 +O(V^3)]. 
\ee
It is thus smaller than the total condensate density, eq.~\eqref{nc}, by a factor that is determined by the glassy fraction \cite{Yukalov2007}
\be 
V_2: = \sum_\bp \frac{\avg{|V_\bp|^2}}{(\epn{\bp}+2g\nc)^2} 
\ee 
with $\epn{\bp}=  \bp^2/2m$ the free dispersion. Furthermore, using eqs.~(18)--(20) of \cite{Mueller2012}, the single-particle
Green's function can be expressed in terms of quasiparticle normal and
anomalous Green's functions, 
\begin{eqnarray}
\avg{G(\bk,0)} =
\sum_{\bp,\bq}
& \Big[\avg{(u_{\bk\bp}u^*_{\bk\bq}+v_{\bk,-\bp}v^*_{\bk,-\bq})G_{\bp\bq}(0)}
\nonumber \\
& \quad - \avg{(v_{\bk\bp}u^*_{\bk\bq} + u_{\bk,
  -\bp}v^*_{\bk,-\bq})F_{\bp\bq}(0)} \Big].  \label{GGF}
\end{eqnarray}
The matrix coefficients $u_{\bk\bp}$ and $v_{\bk\bp}$ generalize the
usual Bogoliubov factors $u_\bk,v_\bk$ to the case where the
condensate, or Bogoliubov vacuum, is inhomogeneous. They encode the
condensate deformation by the external potential $V(\br)$ on the
mean-field level. 
All these factors can be Taylor-expanded to the desired order in $V$
(see Sec.~3.4 in \cite{Mueller2012} and Sec.~III.B. in
\cite{Gaul2011a}). 

To zeroth order in $V$, for the clean
system, one has 
\be\label{Gclean}
\avg{G^{(0)}(\bk,0)} =  -(u_\bk^2+v_\bk^2) \ep{\bk}^{-1} 
\ee
where $\ep{\bk} = [\epn{\bk}(\epn{\bk}+2g\nc)]^{1/2}$ is the
Bogoliubov dispersion.  
Multiplication by $\bk^2$ and taking the limit $ \bk\to0$ as required by \eqref{joseph}
selects the most divergent contribution, which reduces the number of
terms quite substantially. 
\eqref{Gclean} diverges like $  1/(2 a_\bk^2\ep{\bk}) \sim  1/2\epn{\bk} = m/\bk^2$,
such that from \eqref{joseph} one finds 
\be\label{mfclean}
\rhos = m\left|\psi\right|^2 = m\nc =:\rhoc. 
\ee
As expected, in a clean system and to the quadratic order of the Bogoliubov
Hamiltonian considered, the whole condensate is
superfluid. 

At order $V^2$ in disorder strength, two types of contributions
survive in \eqref{GGF}: \\
(i) products like 
\be 
u_{\bk\bp}^{(2)}u_{\bk\bq}^{(0)} G^{(0)}_{\bp\bq}(0)  \propto 
 V_2 \delta_{\bk\bp}\delta_{\bk\bq} u_\bk (u_\bk-2v_\bk)  \ep{\bk}^{-1}
\ee 
with the clean, normal propagator $G_{\bp\bq}^{(0)}(0)=-\delta_{\bp\bq}\ep{\bp}^{-1}$, but no anomalous terms since $F^{(0)} =0$, and \\
(ii) products like 
\be
u_{\bk\bp}^{(0)}u_{\bk\bq}^{(0)} G^{(2)}_{\bp\bq}(0)  \propto 
\delta_{\bk\bp}\delta_{\bk\bq} u_\bk^2 G^{(2)}_\bk(0)  
\ee
and similar with $u_\bk v_\bk F_ \bk^{(2)}(0)$. Mixed terms of the type $u^{(1)}
u^{(0)} G^{(1)}$ and the like do not survive the limit $\bk\to 0$. 

Type (i) terms yield, after taking the limit $\bk\to0$, a
correction $(1-V_2)$ on the right-hand side of \eqref{joseph} that cancels
exactly the same factor introduced on the left hand side by the coherent fraction \eqref{cohfrac}. Type (ii)
terms after a bit of algebra finally yield a correction of the form 
\be
\lim_{\bk\to0}\sum_{\bp}\frac{ (\bk\cdot\bp)^2}{\epn{\bk}\epn{\bp}}
\frac{\avg{|V_\bp|^2}}{(\epn{\bp}+2g\nc)^2}
=  \frac{4m^2}{d} V_2.  
\ee
All in all, \eqref{joseph} predicts to order $V^2$ the correction 
\be\label{rhosV2}
\rhos = \rhoc \left(1-\frac{4}{d}V_2\right), 
\ee 
which is already well documented in the literature, see eq.~(12) in \cite{Huang1992},
eq.~(19) in \cite{Giorgini1994}, eq.~(20) in \cite{Lopatin2002}, and 
eq.~(6) in \cite{Astrakharchik2013}. This then explicitly validates the inhomogeneous Josephson
relation  \eqref{joseph} to order $V^2$ and at the same time rules out \eqref{disorderedJoseph1}. 

Note, though, that one cannot obtain a temperature dependence from \eqref{GGF} with the 
quadratic quasiparticle Hamiltonian of \cite{Gaul2011a,Mueller2012} that contains only elastic impurity scattering. 
In order to recover Landau's celebrated finite-temperature
superfluid depletion \cite{LandauStatPhys2} microscopically, one would have to
introduce interactions between the 
quasiparticles. 


A different method of calculating the superfluid density is to compute
the normal density $\rhon=\rhoc-\rhos$ directly from the transverse current-current
correlation  \cite{LandauStatPhys2,Baym1969}. 
Inhomogeneous Bogoliubov theory \cite{Gaul2011a,Mueller2012} 
then predicts, at $T=0$,  
\be\label{rhon}
\rhon=\frac{1}{4\nc}\sum_{\bp,\bq} \left.
\frac{p_z q_z}{a_\bp a_\bq} \avg{{\nc}_{\bk-\bp}{\nc}_{\bq-\bk} [F_{\bp\bq}(0) -
G_{\bp\bq}(0)]}\right|_{\bk=\bk_\perp\to 0}. 
\ee
Here, ${\nc}_\bk = L^{-d}\int \rmd r e^{-i\bk\br} n_c(\br)$ are the
Fourier components of the deformed condensate density, and $\bk_\perp$ lies in the $xy$ plane transverse to the
$z$ axis. In the clean case, to zeroth order in $V$, the condensate is
homogeneous, ${\nc}_\bq=\nc\delta_{\bq,0}$, and since $k_z=0$, the
normal density vanishes. To order $V^2$,  only a single type of term survives the limit
$\bk_\perp\to 0$, namely 
$n^{(1)}n^{(1)}G^{(0)}$. Using eq.~(11) of \cite{Gaul2011a}, this
expression evaluates 
rather immediately to $(4/d)\rhoc V_2$ and thus agrees with
\eqref{rhosV2}. Clearly, to this order it is much simpler to evaluate
\eqref{rhon} than to find $\rhos$ from the Josephson relation, since there are no common terms that cancel, like on the two
sides of \eqref{joseph}, and only the clean quasiparticle propagator $G_{\bp\bq}^{(0)}(0)$
enters together with the condensate deformation. Lastly, we remark that 
this approach can be generalized to finite temperature and thus permits to 
derive disorder corrections to Landau's superfluid depletion \cite{Mueller_sftocome}.

\paragraph{Summary:---} 

A Josephson-type relation has been established for disordered Bose fluids 
between the superfluid density, the infrared residue
of the single-particle Green's function and the coherent
density, i.e., density of condensed particles with zero momentum. 
Its validity for weak interactions and disorder has been
checked in detail by a perturbative calculation using inhomogeneous Bogoliubov
theory. The numerical results of \cite{Pilati2009,Carleo2013} agree qualitatively
with its prediction at the superfluid-insulator transition where 
coherent and superfluid fraction vanish together.  
Although it may not be evident to extract the
infrared residue of the
average zero-frequency Green
function with precision in the numerics,  it would be interesting to 
investigate the quantitative validity of the sum rule \eqref{joseph} near the
critical point in different dimensions.    

The author is indebted to K.~Krutitsky, S.~Pilati, and M.~Ueda for helpful discussions and to C.~Gaul for constructive comments on the manuscript. 

\bibliography{../../projects}

\end{document}